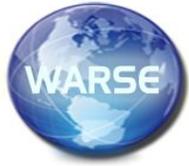

# Blockchain-based System Evaluation: The Effectiveness of Blockchain on E-Procurements

August Thio-ac[1], Alfred Keanu Serut[2], Rayn Louise Torrejos[3], Keenan Dave Rivo[4], Jessica Velasco[5]
[1,2,3,4,5]Electronics Engineering Department, Technological University of the Philippines, Philippines

**ABSTRACT**

Electronic systems tend to simplify the tedious traditional scheme and basically focuses on the platform design and process organization. The integrity of the output of an automated system is not left behind but the possibility of internal manipulation is still high. This paper presents the current issues in company procurements and the solution in the form of blockchain technology. Several individuals and professionals were asked to evaluate a blockchain-based procurement system in comparison to the current electronic (e-procurement) system. A blockchain-based system has the capability to hold transactional data with complete decentralization and eliminate the growing number of fraud cases in companies and organizations. This paper mainly focuses on the effectiveness of a blockchain-based system in company procurements.

## 1. INTRODUCTION

Electronic procurements made a significant change in how companies source goods and services, achieving a better and more cost-effective process [1]. While mainly focusing on process efficiency and cost reduction, fraud and corruption is still present in the system [8]. In the past 2 years, blockchain adoption by different organizations have accelerated, revealing different applications and unlimited possibilities. Blockchain promotes transaction immutability by applying several components that will be discussed in the next section of this text. This study focuses on the application and evaluation of blockchain technology in an online procurement platform. Several private individuals and government organizations were surveyed regarding the effectiveness of a blockchain-based system in the purchasing industry.

## 2. BLOCKCHAIN COMPONENTS

Blockchain network provides a "trust-less" environment, transactions do not rely on a centralized authority [2]. Blockchain provides a kind of write-only data for storing transactions replicated between peers.

Technically, blockchain is a data structure, it is composed of a list of blocks, where each block contains a small or possibly empty list of transactions. Each block in a blockchain is "chained" back to the previous block, that is, by containing a hash of the previous block [3]. With this scheme of ordering transactional data, the blockchain cannot be manipulated without invalidating the previous chain of hashes.

### 2.1 Digital Assets

In addition to the computational constraints, incentivizing the participants on the network in the form of cryptocurrencies, encourages more users to strengthen the consensus mechanism. The technology also has the capability to embed a very small data in the transactions for other useful purposes, such as representing digital assets (e.g., document notarization) [3]-[7]. As shown in Figure 1, blockchain notarization (a useful tool in ensuring digital integrity) is done by obtaining the fingerprint of the digital asset (e.g., hashed file checksum) and recording it on the blockchain. The digital asset fingerprint can be checked on the blockchain to ensure that the information is legitimate and untampered.

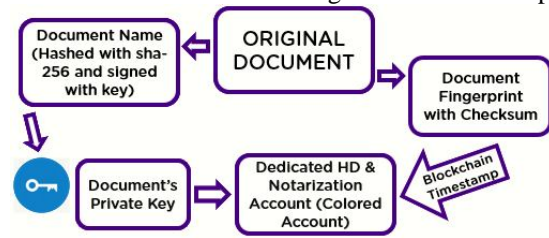

**Figure 1:** Blockchain notarization process

### 2.2 Digital Signature

A user on the blockchain owns a private key and a public key of a respective address (blockchain wallet). The primary role of the private key is for signing transactions. A signed transaction is distributed and recorded on the network. Digital signing is composed of two parts, the signing and the verification. When signing a transaction, the private key is used to encrypt the data. Verifying a message is checking the received data using the public key of the sender, in this way the integrity of the received data can be verified [8].

### 2.3 Multisignature Protocol

The multisignature protocol in the cryptographic world enables digital signing of multiple parties by using their respective private keys to an agreed transaction. This form of digital agreement is useful in cases of a joint financial account. In the case of an outgoing transaction, the participating entities must give their approval by signing the transaction using their private key for the transaction to be processed. In a case of an incoming transaction, the system does not require any action from the participants. In addition, the identity of the participants is also protected [9].

The concept of multisignature protocol is applicable to M-to-N blockchain addresses. Where M is the minimum number of required signers for a transaction to be executed and N denotes the total number of co-signatories. The figure below shows a basic 2-to-3 multisignature account, with Bob, Alice, and Carol as the co-signatories.

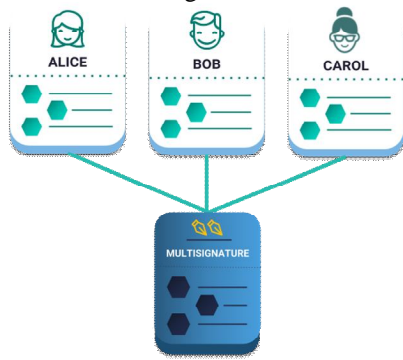

**Figure 2:** 2-to-3 multisignature account as illustrated in [9]

Multisignature accounts can have other multisignature accounts as co-signatories. Multi-level multisignature accounts enables "AND/OR" logic to blockchain transactions [10]. The usefulness of multisignature protocol can be realized in this type of transaction, this illustrates the flexibility of this blockchain feature.

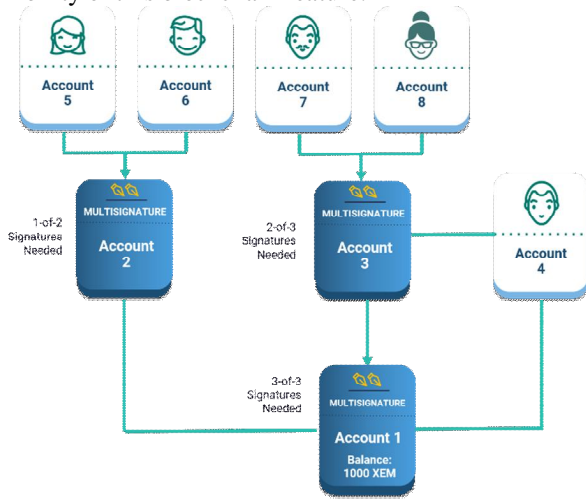

**Figure 3:** Three-level multisignature account example as illustrated in [10]

### 2.4 Smart Contracts

Second generation blockchains enables a programmable infrastructure in the form of smart contracts. Smart contracts are programs that can be deployed and run on top of blockchain to enable automation of complex transactions. It can be triggered when certain conditions are met [7]. Blockchain smart contracts have the capacity to disrupt various fields in financial services and Internet of Things (IoT) [4]. Internet of Things can communicate the accessible input data resources and probable output applications e.g. medical, energy, automation together through the Internet [11]-[15].

### 3. PROCUREMENT FRAUD

Procurement fraud is the third leading economic crime in the Philippines based on the 2018 Global Economic Crime and Fraud Survey conducted by PwC [16]. Asset misappropriation is at 53%, followed by business misconduct at 38% and procurement fraud at 35%. Based on the survey 98% of the 63 respondents indicated that they have knowledge of fraud incidents in their organization. The organizational roles of the respondents are seen on the graph below.

This data suggests that the amount of the reported fraud in the organization is concentrated in the upper hierarchy. This type of internal issue cannot be easily resolved and detected. Even though electronic systems can promote efficiency and cost reduction [1], the possibility of internal manipulation by the governing authority is still present in the picture.

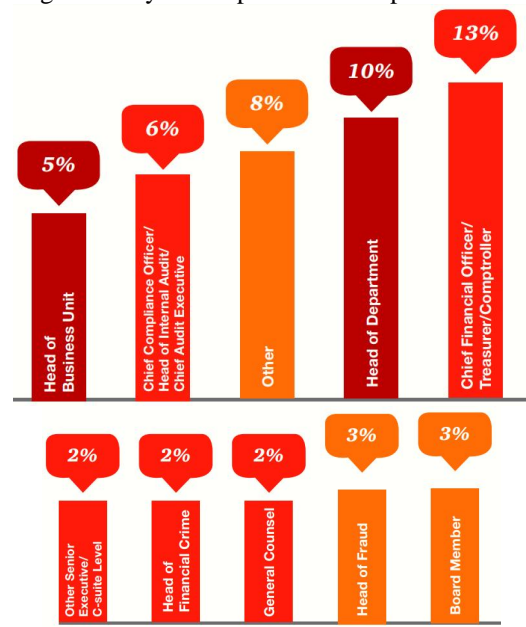

**Figure 4:** Job roles of PwC survey respondents as illustrated in [16].

### 4. BLOCKCHAIN-BASED SYSTEM

Blockchain-based systems are equipped with the blockchain components discussed in the previous section of this text. Blockchain-based systems promote data immutability and even more cost reduction when applied in electronic systems. Cost efficiency is achieved by cutting out intermediaries while achieving a high level of security [3]. Blockchain promotes a "trust-less" environment where transactions are not governed by a central authority, promoting complete decentralization [2].

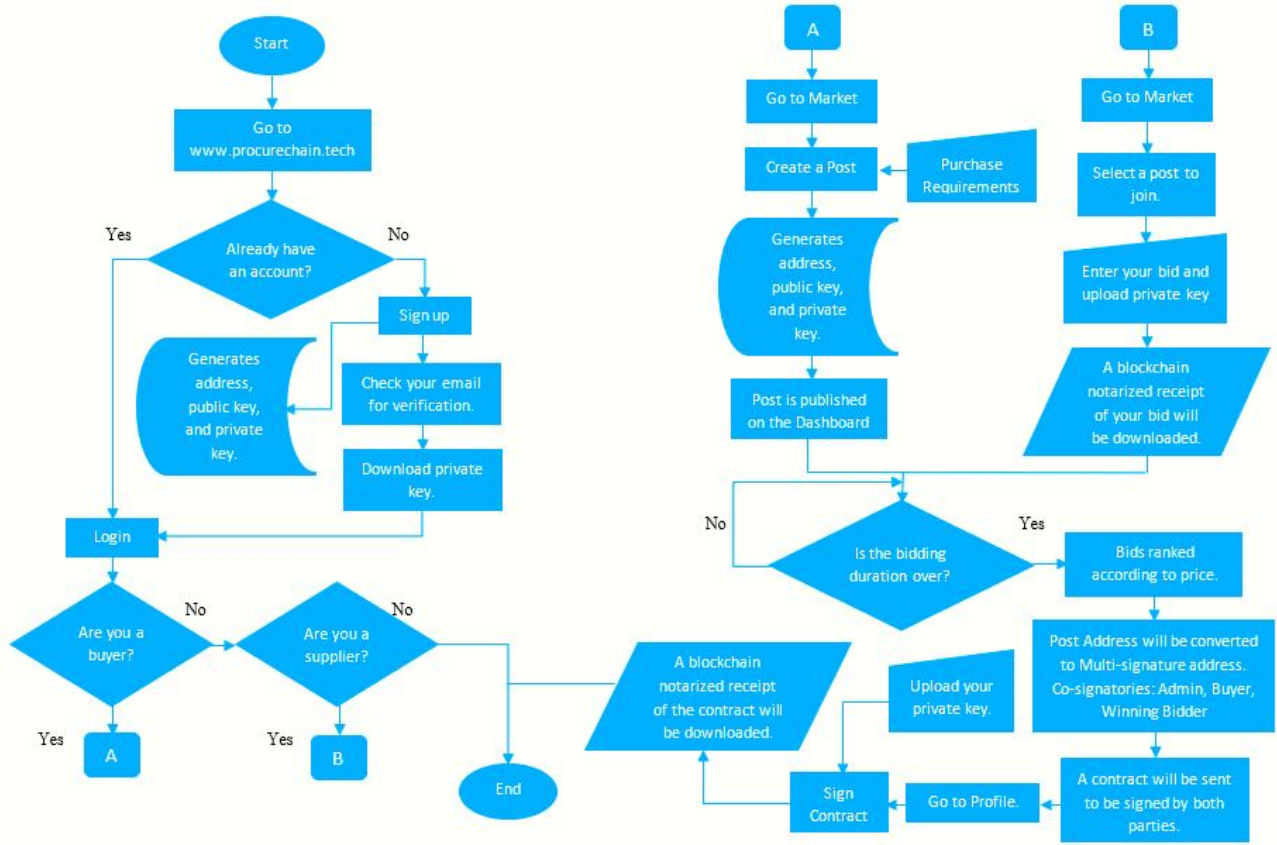

**Figure 5:** A process model of a user joining the platform and acting as a buyer and a supplier.

## 5. METHODOLOGY

A blockchain based procurement platform was developed using three blockchain components – digital signature, multisignature protocol, and blockchain notarization. The blockchain platform only focuses on blockchain integration in a procurement process. The program is limited to the ranking of supplier's bid according to price. A sample flow process in the platform is shown above in Figure 5.

Several private entities were asked to use the platform, focusing on its key blockchain features. An evaluation sheet is used to collect the response of the users based on their experience using the platform. In addition to the collection of data from private individuals, an interview with four government procurement related entities were conducted.

Specifically, the evaluation sheet is composed of 10 key parameters focusing on the efficiency and level of security from using blockchain technology, the responsiveness of the platform, and the user experience.

## 6. DATA AND RESULTS

To measure the effectiveness of the blockchain-based online procurement system, a survey was developed to collect empirical data from verified users of the site. Specifically, the survey was structured to evaluate the site with the following variables: 1) Efficiency, 2) Responsiveness, 3) Safety and Security, and 4) Aesthetics. The survey was circulated to a total of 64 respondents and utilized the five-level Likert Scale. In terms of efficiency, 54.89% of the respondents strongly agrees, 39.06% agrees, and 6.25% are undecided that using an electronic and automated system from purchase posting up to the issuance of contract is efficient. Meanwhile, 71.87% of the respondents strongly agrees, 25% agrees, and 3.13% are undecided that time is saved by using the online platform. 71.87% of the respondents strongly agrees, 18.75% agrees, and 9.38% are undecided that the bidding processes were conducted correctly and according to the indicated timeframe. In terms of safety and security, the result shows that 82.81% of the respondents strongly agrees, 4.61% agrees, and 12.5% are undecided that blockchain notarization features for a tamper-proof contract and bid receipt is useful in securing digital documents. 67.19% of the respondents strongly agrees, 23.43% agrees, and 9.38% are undecided that the file auditing feature for checking the legitimacy of the file is effective. 71.87% of the respondents strongly agrees, 25% agrees, and 3.13% are undecided that transactions conducted are completely immutable. 84.37% of the respondents strongly agrees, and 15.63% agrees that the presence of a Know-Your-Customer verification feature for checking the identity of the user is an effective way of ensuring that the participants of the platform is legitimate.

## 7. CONCLUSION

In this paper, blockchain integration on a simple electronic procurement platform is introduced, utilizing blockchain notarization, digital signature, and multisignature protocol for better security while achieving cost-efficiency.

While many studies have already proven the effectiveness of the disruptive technology in different domains of application, the possibility of public usage of this type of system is still miles away from happening. Several private organizations in other countries have also applied blockchain technology in different areas like in supply-chain management. This study is a step forward to achieving a trust-less environment showing how a paperless transaction with the use of blockchain technology can guarantee a better security.

**ACKNOWLEDGEMENT**

The proponents of this study would like to thank the University Research and Development Services Office of the Technological University of the Philippines for the financial support. They would like to acknowledge Ms. E. M. Paule for her voluntary services such as editing and proofread of the paper.